\documentclass[twocolumn]{article}
\usepackage{graphicx} 
\usepackage{hyperref}

\title{The Role of Likes: How Online Feedback Impacts Users' Mental Health}
\author{
    Angelina Voggenreiter\textsuperscript{*}\textsuperscript{1}\\
    0000-0001-6597-3514\\
  \and
    Sophie Brandt\textsuperscript{*}\\
    0009-0003-6999-5106
    \and
    Fabian Putterer\textsuperscript{*}\\
    0009-0009-9310-3634
    \and
    Andreas Frings\textsuperscript{*}\\
    0009-0009-4993-0384
    \and
    J\"{u}rgen Pfeffer\textsuperscript{*}\\
    0000-0002-1677-150X\\
}

\date{\textsuperscript{*}School of Social Sciences and Technology, Technical University of Munich \\ \textsuperscript{1}angelina.voggenreiter@tum.de}
\begin{document}

\maketitle

\begin{abstract}
Social media usage has been shown to have both positive and negative consequences for users' mental health. Several studies indicated that peer feedback plays an important role in the relationship between social media use and mental health. 
In this research, we analyse the impact of receiving online feedback on users' emotional experience, social connectedness and self-esteem. In an experimental study, we let users interact with others on a Facebook-like system over the course of a week while controlling for the amount of positive reactions they receive from their peers. We find that experiencing little to no reaction from others does not only elicit negative emotions and stress amongst users, but also induces low levels of self-esteem. In contrast, receiving much positive online feedback, evokes feelings of social connectedness and reduces overall loneliness. On a societal level, our study can help to better understand the mechanisms through which social media use impacts mental health in a positive or negative way. On a methodological level, we provide a new open-source tool for designing and conducting social media experiments.
\end{abstract}


\section{Introduction}
Worldwide, more than 4.5 billion people use social media platforms, such as Facebook, Twitter or Instagram. This number is expected to rise in the following years, with an estimated 5.85 billion users by 2027 \cite{Statista1}. Based on the sheer amount of people actively participating in such online spaces, it is important to understand the impact social media can have on their users. 

On the one hand, social media offers tremendous possibilities to improve people's mental health: Social media has been shown to provide distinct social support and connectedness, which can be associated with better mental health outcomes \cite{Seabrook_Kern_Rickard_2016}. 

Especially when looking at marginalized groups such as the LGBTQIA+ community, social media can serve as a place to find friends and social support. For example, LGBTQIA+ youth is more likely than non-LGBTQIA+ youth to report having friends exclusively known online, who are further described as more supportive than their in-person friendships, proving how the internet can serve as a safe space for receiving both social and emotional encouragement \cite{Ybarra_Mitchell_Palmer_Reisner_2015}. By investigating the online behaviour of university students an overall positive indirect impact of social media use on psychological well-being was found. The effect was mainly derived from its positive effect on bonding and bridging social capital, implying that the use of social media allows students to continue their close relationships during events such as a pandemic. Further, social media was shown to improve trust and promote the establishing of social connections. Finally, it was found that students can use social media platforms to receive emotional support and increase their potential to mobilize others or build social networks, leading to social belongingness \cite{Ostic}. 

On the other hand, social media use has also been associated with numerous mental health issues, such as anxiety, depression, loneliness, poor quality of sleep, thoughts of self-harm and suicide, and increased levels of psychological distress \cite{Sadagheyani2021}. 

Coyne et al. \cite{Coyne2020} performed an eight-year longitudinal study to analyse the effects of social media use on adolescents' mental health. Based on annual surveys, they found that, although there were no within-subjects relationships between social media usage time and mental health consequences, participants who spent more time on social media had a higher chance of experiencing anxiety and depression, compared to others \cite{Coyne2020}.

Further investigation into social networks and their relationship to student mental health showed that symptoms of poor mental health, especially depression, increased with a roll-out of Facebook at the college. For students predicted to be most susceptible to mental illness, the introduction of Facebook led to an increased uptake of mental healthcare services. Additionally, the likelihood to report a negative impact of poor mental health on their academic performance increased \cite{Braghieri2022}. Another study looking specifically at adolescents showed higher usage of social media and higher emotional investment in social media leads to a poorer quality of sleep, higher levels of anxiety and depression, and lower self-esteem \cite{Woods2016}. 

However, the mechanisms in which social media might impact mental health in a positive or negative way are highly complex. To develop better social media platforms and help people have positive online experiences, we need to understand which factors exactly determine the emotional state of the user when engaging with social media. In this study, we aim to shed further light on this question and make the following contributions: 
\begin{itemize}
    \item We show that repeated absence of feedback in Social Media elicits negative emotions, such as stress and anxiety, as well as low levels of self-esteem.
    \item We demonstrate that a positive online experience, characterized by receiving many \textit{likes}, evokes feelings of social connectedness and reduces overall loneliness.
    \item We provide an open-source tool for social science researchers to design and conduct experimental studies in a Facebook-like environment and collect data about user behavior, without needing any programming skills or technical resources.
\end{itemize}

\section{Related Work}

\subsection{The Relevance of Peer Feedback}
Several studies indicated that the negative association between social media use and mental health might be connected to peer feedback \cite{Radovic2017, Gallagher2017, Diefenbach}. Many social media platforms, like Facebook and Instagram, allow participants to react to each others content, for example by distributing social affordances, such as \textit{likes} or hearts. Often, statistical metrics on the amount of reactions, such as the \textit{like} or follower count, are calculated and displayed prominently on the platform. Thus, users are not only confronted with the number of feedback they have received by others, but also with the number of feedback other users have received by their peers, allowing for comparisons in regards to online popularity. 

When researching social media use and its impact on adolescents diagnosed with depression, Radoviv et al. \cite{Radovic2017} found that peer feedback might play a prominent role in the connection between social media consumption and mental health. Within their qualitative interview study, they tried to understand how psychological distress may affect social media use or how this usage may influence psychological distress. They found that social media are employed as a measure to gain social approval and acceptance as well as to compare oneself with others. According to their results, this manifests specifically by checking the \textit{like} count, which is conceived as a unit of comparison and turned into an indicator of popularity. Consequently, receiving more \textit{likes} seemed to increase self-esteem, while not receiving \textit{likes} decreases it. Even more, adolescents seemed to become motivated to create posts they feel less comfortable with, such as suggestive pictures, to achieve the craved recognition \cite{Radovic2017}.  

In line with this stand the results of Gallagher \cite{Gallagher2017} who analyzed the relationship between social media use and teenagers' self-esteem. Based on two surveys, she identified four variables which influenced teenagers' self-confidence. Interestingly, all four variables were more or less directly associated with \textit{likes}. In particular, she identified that self-esteem was influenced not only by the amount of \textit{likes} on a participant's last selfie or the amount of \textit{likes} they would usually receive, but also the time between posting on the platform and checking on the post. In addition, not receiving as many \textit{likes} on a post as the participant anticipated, seemed to decrease self-esteem.

These results were related to the study of Diefenbach and Anders \cite{Diefenbach} who used a survey to investigate the subjective relevance of feedback on Instagram, which can be understood as the individual differences concerning the importance ascribed to other users' feedback by different individuals. Self-esteem was revealed to be negatively correlated with the subjective relevance of feedback quantity and quality. Therefore, putting high importance to receiving many as well as positive reactions from others on Social Media was associated with a lower self-esteem.  Finally, they concluded their findings by noting that feedback relevance predicts users' psychological states more than specific activities or events on Instagram. 

On a macro level, there seems to be a connection between feedback on social media, psychological distress and self-esteem. However, the complex patterns of social media usage and the manifold ways, in which one could receive peer feedback online, make it hard to uncover the underlying mechanisms within this relationship. For example, based on these studies, it remains unclear whether social media feedback, or the lack thereof, only impact users' emotional experience and self-esteem over a longer period of time or whether the underlying mechanisms are directly at work. 

\subsection{Micro-Level effects of Social Media Ostracism}
To gain more insights into these relationships, several studies have researched these connections on a micro-level. 

In their fMRI-Study, Wikman et al. \cite{Wikman2022}, asked adolescents to respond to a controversial statement by either agreeing or disagreeing. Their response was then presented as a social media post together with four comments which were all either positive or negative. The task was repeated multiple times, allowing participants to read the reactions of social media posts over the duration of an hour. During this task the participants brain activity (fMRI) was measured, showing that negative comments activated different brain regions than positive feedback. For example, negative social media feedback lead to an increased activation of regions related to emotion regulation, indicating that users experienced negative emotions due to peer feedback. Thus, the study showed that, whatever happens to people during receiving online feedback, does so directly. 

Similarly, multiple researchers analyzed the direct effects of social media feedback on users' emotional experience and self-esteem. They showed that even very short interactions of online ostracism, that is being socially excluded or ignored in a virtual setting, can lead to negative emotions as well as lower short-term self-esteem.

For example, Wolf et al. \cite{Wolf2014} manipulated the number of virtual \textit{likes} participants received from others and analyzed the effects on emotions, feelings of having a meaningful life, self-esteem, and others. In their study, participants were asked to write a short self-description and read through the online description of other "group members" (actually generated by computer scripts). They should then like the descriptions of others and could see the number of \textit{likes} they and their group members received. Depending on the treatment condition, participants either received few, medium or many \textit{likes} in comparison to their group. The study showed that participants receiving little attention from others reported more negative and less positive emotions, lower feelings of having a meaningful existence and lower levels of self-esteem than participants receiving medium or many \textit{likes}. 

To study the effects of online ostracism on adolescents' situational emotions and self-esteem, Lee et al. \cite{Lee2020} repeated the Wolf-Task, with two conditions (few vs. many \textit{likes}) and a ranking board showing the level of \textit{likes} one received in comparison to others. Similarly, participants receiving fewer \textit{likes}, reported  having experienced more negative emotions, feelings of rejection and lower self-esteem during the task. In addition, they found that the effect of feeling rejected when receiving few \textit{likes} in comparison to many, was higher for students who were victimized by peers in the two weeks prior to the social media task.

While these studies indicated that the number of virtual \textit{likes} can affect people's emotions and self-esteem, the study setup consisted of a very short time intervention, which might not reflect the full spectrum of social media interactions. Specifically, participants could get to know their social media partners only by reading, solely with one description about them. Crucially, they only received \textit{likes} based on a single short post and they could see their \textit{likes} only for a few minutes. In our study, we wanted to analyze the effects of social inclusion and exclusion in a setting which would, to a larger extent, resemble real-life social media experiences of users. Thus, we did not limit the interaction to a couple of minutes, singular posts or a snapshot in time but rather placed it over the duration of one week with daily engagement on the platform. This should allow participants to socially connect to their peer group by reading about their lives, experiences and emotions on a daily basis during the study. Additionally, unlike previous studies where receiving few (many) \textit{likes} was a one-off experience, we implemented social exclusion (inclusion) as a repeated phenomenon for every post shared on the platform. Further, the environment in which the study took place closely resembled the already existing social media platform Facebook. By using this social media design we investigated the following question:

\begin{quote}
RQ1: Does repeated social media feedback (or the lack thereof) impact users' emotions, feelings of social connectedness and self-esteem during social media interaction?
\end{quote}

Furthermore, as Desjarlais \cite{Desjarlais2020} describes "All of research in this area has examined momentary fluctuations in state self-esteem, and thus is unknown whether these changes have more long-term effects." To analyze the effects of peer feedback beyond this state self-esteem, we differentiated in our study between context-dependent and context-independent self-esteem, and evaluated changes within stable self-esteem. In particular, we asked:

\begin{quote}
RQ2: Does social media feedback invoke changes within overall, context-independent self-esteem?
\end{quote}

\subsection{Studying Social Media}
In the past, performing experimental studies about behavioural patterns on social media has proven challenging. In order to achieve a high level of ecological validity, it would be preferable to conduct studies in a realistic social media environment. However, researching user behaviour directly on a social media platform, like Facebook, imposes various limitations. For example, scientists usually only have a small influence and control on the content displayed on a social media platform. In addition, the platforms usually cannot be freely adapted by the experimenters and thus researchers may not be able to compare the impact of different functionalities (such as including or not including a functionality to like content). Moreover, studies on social media platforms impose ethical challenges, such as gaining informed consent of social media users, and thus, company regulations often prohibit the conduction of experimental studies on the platform. Lastly, researchers might not have access to all data needed for evaluating behaviour on the platform, even when they collaborate with the platform provider \cite{Wagner2023}. 

To prevent these challenges, a controlled, but still realistic, environment for social media studies is needed. While there already exist great tools allowing researchers to set up and customize such environments \cite{Truman, MockSM}, these environments do not allow to study interaction between participants. In our work, we developed an open-source tool, which allows researchers to design and perform experimental studies in a social media environment, without requiring any programming skills. With this tool, we could influence the content users see and the (inter)actions they can perform, and collect extensive behavioral data.

\section{Study Methodology}
\subsection{Data}
The study was conducted between October 2022 and September 2023. Participants were invited to take part in this research through online advertisements within groups for interested study participants as well as university channels. The participants were compensated with €30 for their time investment of approximately two hours over a duration of one week. In total, 275 people registered for the study. We excluded all participants, who did not fully complete all tasks as described below, including publishing a post a day for a period of five days, interacting with others and completing two questionnaires. 

The resulting dataset size comprised 170 participants (85 per treatment). The participants were between 19 and 59 years old (M = 26.8, SD = 5.7) and were highly educated, with 116 having a university degree and 48 having a qualification for university entrance. 98 of them identified as female, 71 as male, none as 'other' and one person did not specify their gender. All participants used social media at least once a week, except one who used it at least once a month, and 54\% of participants used Facebook at least once a week. 

All data was stored on university resources in a pseudonymized way and each user was assessed with an unique identifier so that researchers could relate pre- and post-measurements.

\subsection{Design}
The study was conducted in German. The participants were told that they would take part in a study on social media and marketing and that they would interact with other participants within a Facebook-like environment for a period of five days. 

\begin{figure*}[t]
   \centering
   \frame{\includegraphics[width=2.0\columnwidth]{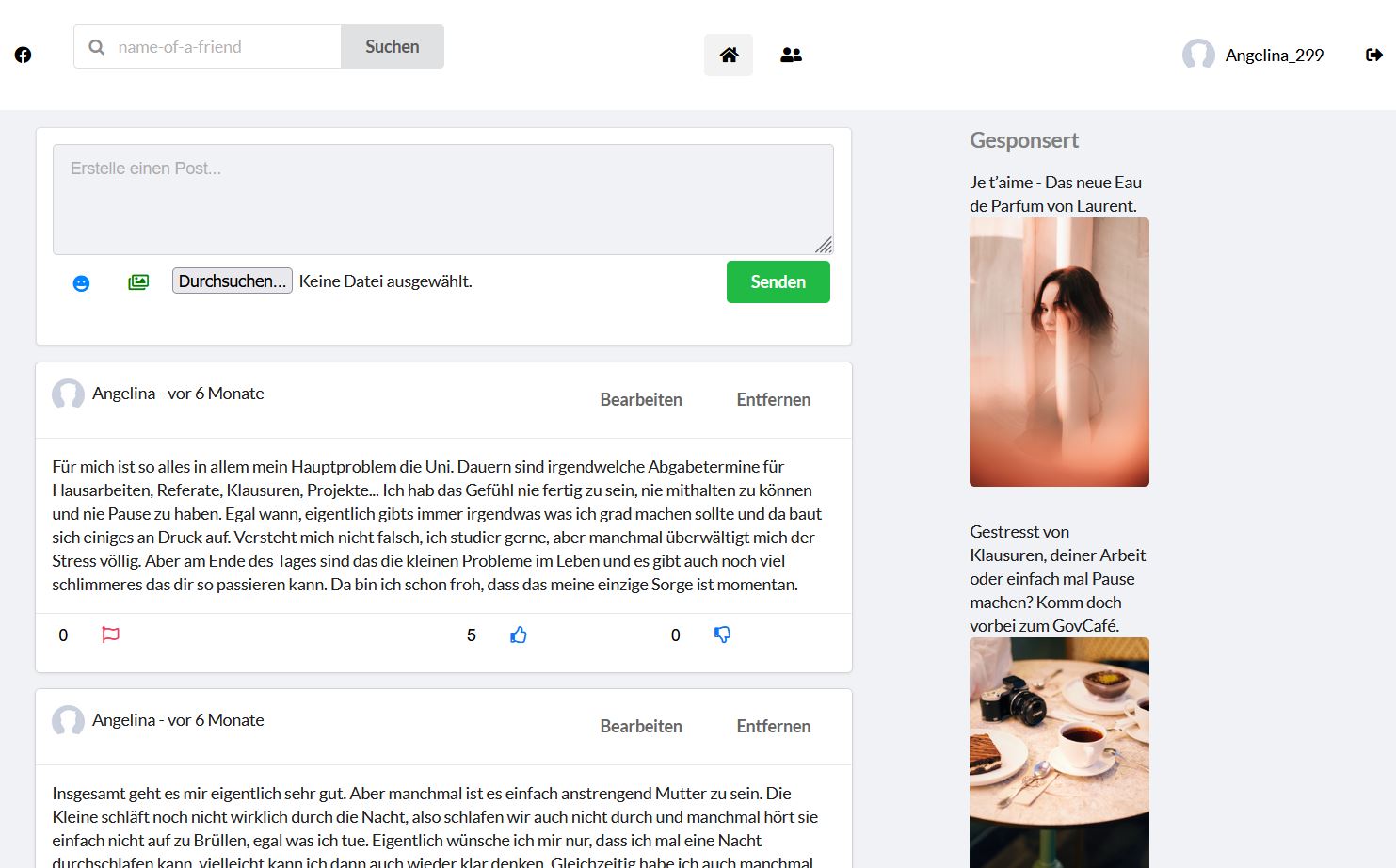}}
   \caption{Fakebook participant interface adapted for the study needs. The screenshot displays two posts (not of a real participant due to privacy concerns) and two advertisements (on the right), as well as functionalities to create posts, like, dislike and flag content. Functionalities to send friend requests and chat messages were disabled for the study and are thus not visible.}
   \label{fig:fb}
\end{figure*}

We then randomly assigned each participant to a treatment condition ("likes condition" vs. "no likes condition") and to their own account on our study platform, with a group of six bots (which they believed were other participants). Each study subject could only see the posts, profiles, and interaction activity of themselves and these bots. Within the system, study participants could create their own posts, react to the posts of bots and see which and how many accounts have liked their posts.  Figure \ref{fig:fb} shows the environment participants interacted within.

Before entering the system for the first time, each participant conducted a survey including a consent form\footnote{To mitigate any potential participant risks, participants were informed that they could withdraw from the study at any time. Moreover, we offered email support to participants on a daily basis, which they only used for technical questions.} as well as questions on demographics, social media use, feelings of loneliness, and self-esteem. 
On day one to five, participants received an email in the morning as well as a reminder email in the evening, with the task to be active on the Fakebook platform for at least 15 minutes, to like the posts of at least two other "participants" (i.e. bots) and to create a post with at least 600 characters on a specific topic. These topics were chosen in a way that study subjects would need some time to interact with their content and emotionally connect to it, e.g. "Publish a post about what you want to leave to the world. What big goals do you have? What are the values that you live by?". 

Within this time frame, participants and bots received a pre-specified number of \textit{likes} on their post from other bots. All bots in both conditions received zero to five \textit{likes} per post, resulting in two bots with a low (two to three), two bots with a medium (12) and two bots with a high (24) sum of \textit{likes} over the week.
As participants believed that the other "participants" should also like the posts of at least two people, they should have received on average at least two \textit{likes} per post. However, we distributed \textit{likes} in a way matching the treatment condition: Study subjects of the \textbf{many likes condition} received \textit{likes} of \textbf{four to five} bots per post, summing up to 24 \textit{likes} within these five days. In contrast, participants in the \textbf{few likes condition} received \textbf{no likes} on their posts, except one \textit{like} on their first post to let them realize that it would be technologically possible for them to receive and see their own \textit{likes} (and thus receiving no \textit{likes} would not be attributed to technical issues). 

\begin{figure}[t]
   \centering
   \includegraphics[width=1.0\columnwidth]{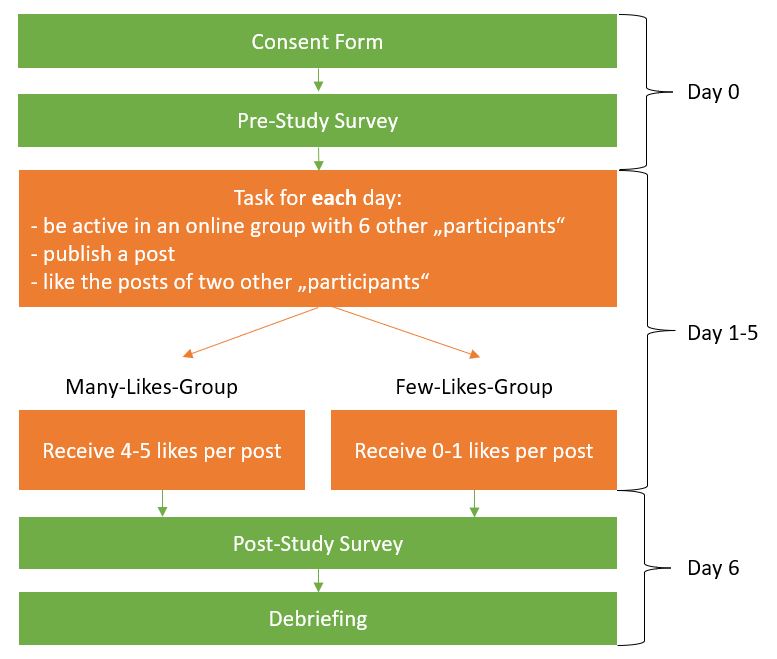}
   \caption{Overview of the study design.}
   \label{fig:des}
\end{figure}

On day six, participants were instructed to be active on the platform for another 10 minutes, e.g. with reading or liking posts. After this, they completed a second questionnaire including questions on situational emotions, social connectedness and self-esteem. Finally, the participants were informed about the aims and design of the study, including the information that they had interacted with bots instead of real users. Figure \ref{fig:des} gives a short overview of the study design.

\subsection{Setup}
To perform this experimental study, we developed a tool named Fakebook \footnote{\label{note:tool}\url{https://github.com/jpfefferlab/Fakebook}}, which would allow us to setup a Facebook-like environment, let participants interact within this environment and collect extensive data on user behaviour by using simple clicks. In the following, we will describe how we used the tool to design and conduct our study.

Using the Fakebook admin panel, we disabled all functionalities to exchange chat messages and comment on posts within the system, as participants should only interact with others by posting and liking content. In addition, we disabled functionalities to send friend requests and enabled the option that users would only see posts by friends, to ensure that participants would only see the content we exposed them to (and not the content of other participants). To make the study environment more realistic, we created two fake advertisements, one about perfume and one about travelling.

Next, we created a set of six bot profiles, which included different genders, age groups, interests and nationalities. For each bot, we created five different posts, one for each day of study, with a timestamp relative to the participant's start date. All posts by bots were designed to be authentic, including different writing styles and use of emojis, descriptions of relatable situations and emotions, and spelling mistakes, but did not include any offensive content. In addition, for each bot we created a set of like actions with a timestamp relative to the posting date, to define at which time a certain post of a bot or the participant should be liked. All profiles, posts, and like actions were saved as csv-file.

When a new participant registered for the study, we created a new participant account and used the upload csv-data function in the admin panel, upload the six bot accounts, planned posts and like actions. We then befriended all accounts with each other, so that the participant would be able to see and interact with content of the bot profiles. We then send the link to the user environment on the platform as well as login credentials to the user account via email to the participant. After login, study participants could interact within the environment, e.g. by reading and creating posts, reacting to content, clicking on advertisements, or inspecting the profiles of bot accounts. 

During the study, the system automatically collected data about user behavior and stored them in a database. This data could be downloaded any time during the study as csv-file and included all information generated on the platform, such as posts and comments published, all reactions to posts and comments (e.g. like, flag,...), all profile descriptions and all messages and friend requests sent between users. In addition, it included user tracking data, such as user log information with session ID, start and end, data about how long each post was viewed by a user and data about when a user has clicked on an advertisement. As soon as a user was finished with the study participation, we downloaded his or her data to inspect whether the participant had completed all of the tasks (e.g. such as publishing a post per day).

\subsection{Measures}
Study participants conducted two surveys, one before and one after the five days of social media interaction.

\subsubsection{Situational emotions.}
Emotions during the social media interaction were measured with single items in the post survey, which included enjoyment ("I enjoyed the interaction on Fakebook."), stress ("I felt stressed during the interaction on Fakebook"), anxiety ("I felt nervous during the interaction on Fakebook") and sadness ("I felt sad during the interaction on Fakebook"). Participants could respond to each statement on a 5-point-scale (-2 = strongly disagree to 2 = strongly agree). 

\subsubsection{Social connectedness.}
Feelings of social connectedness during the social media interaction were measured with single items in the post survey, which included feelings of belongingness ("During the social media interaction, I felt well received by the other participants."), appraisal ("During the social media interaction, I felt valued."), and rejection ("During the social media interaction, I felt rejected by the other participants"). Participants could respond to each statement on a 5-point-scale (-2 = strongly disagree to 2 = strongly agree). 

In addition, a measure of loneliness was included both in the pre- and post-survey, to analyze potential changes in how lonely people feel after five days of social media interaction in comparison to before. To measure loneliness, we used a German short-version of the UCLA loneliness scale, consisting of 10 items. For each item (e.g. "I feel excluded"), participants reported the frequency to which the item applied to their lives, using a 5-point scale (1=never to 5=very often). After this, an overall loneliness score was calculated, ranging from 10 (= not lonely) to 50 (= extremely lonely).

\subsubsection{Self-esteem.} 
To analyze the effects of social media feedback on self-esteem, we used two different concepts: As context specific measure of self-esteem, we quantified the self-referent attribution during the social media interaction, which we will define in the following as “situational self-esteem”. This was measured with a single item: “During the interaction on Fakebook, I felt like a person of worth, at least on an equal basis with others.” (-2 = strongly disagree to 2 = strongly agree) in the post-survey. 
In contrast to that, we measured the rather context-independent self-esteem, in the following named as “stable self-esteem”, by using the German version of the 10-item Rosenberg Self-Esteem Scale \cite{vonCollani2003}. Participants reported the extent to which they agreed with each item (i.e. “On the whole, I am satisfied with myself”) using a 4-point scale ranging from 1 (strongly disagree) to 4 (strongly agree). The scale was included in both the pre- and post-survey to allow for measurements of potential change. After this, an overall self-esteem score was calculated, ranging from 10 (= no self-esteem) to 40 (= high self-esteem).

\subsubsection{Reasoning.}
To assess how people would attribute getting few or many \textit{likes} during the social media interaction, we let study participants estimate the amount of \textit{likes} they had received as well as the amount of \textit{likes} the other participants had received and then posed the open question "Why do you think, did you get the amount of \textit{likes} you received during the social media interaction?". The answers of seven participants were excluded, as they were incomplete or did not refer to the question. We then openly coded the reasons stated, resulting in five categories with reasons related to post content, to person, to posting time, to study design or no reason given. In case participants stated multiple reasons, they were counted once per category.

\subsection{Analysis}
Making use of a between-subject design, we compared the central tendencies of the levels of enjoyment (stress / anxiety / sadness / feelings of belongingness / feelings of appraisal / feelings of rejection /situational self-esteem) in the many likes condition with the levels of enjoyment (stress / anxiety / sadness / feelings of belongingness / feelings of appraisal / feelings of rejection /situational self-esteem) in the few likes condition. As the attributes were ordinal scaled, we used a two-sided Mann-Withney U test, against the null hypothesis that there would be no differences in the central tendencys between the two groups.

In addition, utilizing a within-subject design, for each group we compared the central tendencies of loneliness (stable self-esteem) before the social media interaction with those after the social media interaction. Due to violations of the assumption of normal distribution and as we compared two dependent samples, we used for each group a two-sided Wilcoxon signed rank test, against the null hypothesis that there would be no differences in the central tendencys before and after the interaction.

\subsection{Availability and reproducibility}
To allow for further replications of the study, all steps of the study and accompanying study materials, including advertisement for the study, full text of instructions given within the 13 emails, consent form, surveys, and bot posts were carefully documented and will be provided on request (see email address in paper head).

Even more, to allow for performing other social media research, we offer access to the social media tool in two different ways\footref{note:tool}: Researchers with less technical expertise can get free online access to the tool, including that data is stored on a university server within the EU, and the researcher receives sole\footnote{To ensure privacy, we as tool provider can add or delete a new instance for researchers, but will not be able to inspect any data created within the instance.} admin access for this instance, with which he or she can inspect and control all data on this instance. Researchers with more technical expertise can download the open-source project from a Github repository and either install the system on their own server with a Docker container or with manual deployment. 


\section{Results}
\begin{table*}[t]
\begin{center}
\begin{tabular}{ ccccc} 
 & Many Likes Condition & Few Likes Condition & \\
 & M(SD) & M(SD) & Group Difference\\ 
  \hline
Stress  	& -.92 (1.06) & -.28 (1.16) & \textit{U\textsubscript{min}}=2477.0, \textit{r}=-.31, \textit{p}$<$.01* \\
Sadness  	& -1.2 (.88) & -.59 (1.22) & \textit{U\textsubscript{min}}=2590.0, \textit{r}=-.28, \textit{p}$<$.01* \\
Anxiety  	& -1.18 (1.07) & -.68 (1.16) & \textit{U\textsubscript{min}}=2718.5, \textit{r}=-.25, \textit{p}$<$.01* \\
Enjoyment  	& .59 (.99) & -.25 (.91) & \textit{U\textsubscript{min}}=1959.5, \textit{r}=.46, \textit{p}$<$.01* \\
Belongningness  	& .99 (.84) & -.85 (.91) & \textit{U\textsubscript{min}}=596.5, \textit{r}=.83, \textit{p}$<$.01* \\
Appraisal  	& .66 (1.04) & -.65 (1.01) & \textit{U\textsubscript{min}}=1392.0, \textit{r}=.61, \textit{p}$<$.01* \\
Rejection  	& -1.46 (.8) & .32 (1.13) & \textit{U\textsubscript{min}}=851.0, \textit{r}=-.76, \textit{p}$<$.01* \\
Sit. Self-Esteem  	& 1.19 (.84) & .51 (1.17) & \textit{U\textsubscript{min}}=2426.5, \textit{r}=.33, \textit{p}$<$.01* \\
\hline
\end{tabular}
\end{center}

\caption{Descriptive statistics and differences between users who received many \textit{likes} and users who received few \textit{likes}. The \textit{U\textsubscript{min}}-values represent the test statistics for the Mann-Whitney U tests and the \textit{r}-values express the rank-biserial correlations as measure of effect size.}
\label{tab:manwhu}
\end{table*}

\subsection{Situational emotions}
\begin{figure*}[t]
   \centering
   \includegraphics[width=2.0\columnwidth]{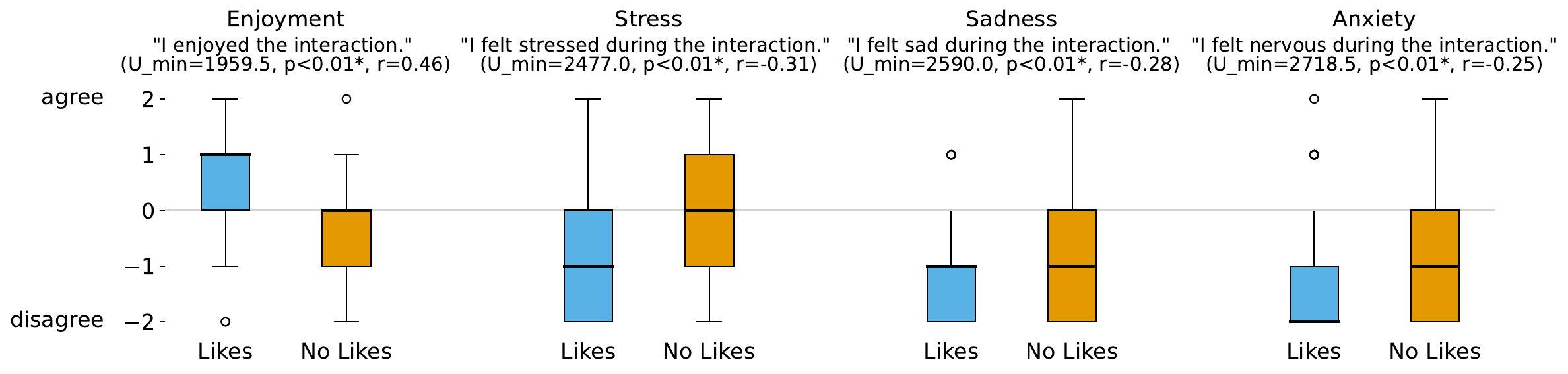}
   \caption{Boxplot visualizing the distributions of situational emotions in the many \textit{likes} and few \textit{likes} condition. Thick black lines indicate the median values per groups.}
   \label{fig:emotions}
\end{figure*}

When asked about their feelings during the social media interaction on Fakebook, participants who received few \textit{likes} rated themselves significantly higher according to the extent to which they experienced stress, sadness and anxiety than participants in the many \textit{likes} condition. Even more, they reported enjoying the interaction to a much smaller extent than participants receiving many \textit{likes}. The results of the statistical tests are shown in table \ref{tab:manwhu} and the differences are visualized in figure \ref{fig:emotions}.

\begin{figure*}[t]
   \centering
   \includegraphics[width=2.0\columnwidth]{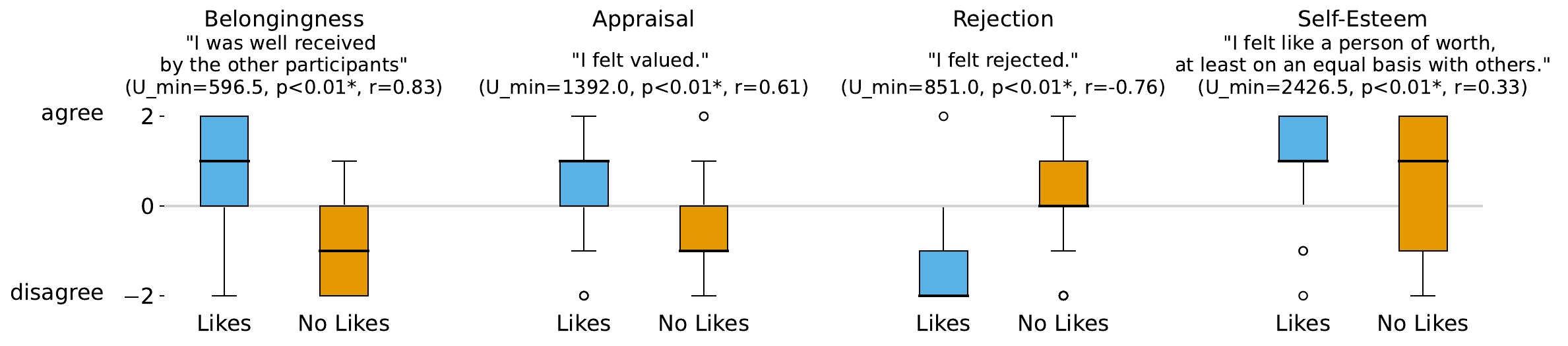}
   \caption{Boxplot visualizing the distributions of social connectedness and situational self-esteem in the many \textit{likes} and few \textit{likes} condition. Thick black lines indicate the median values per groups.}
   \label{fig:selfesteem}
\end{figure*}

\subsection{Social connectedness}
Whether participants received \textit{likes} or not, had significant effects on how connected they felt to others during the social media interaction on our Fakebook.
In comparison to participants receiving many \textit{likes}, participants who received few \textit{likes} seemed to feel less welcomed and less valued by their peers during the interaction. Furthermore, participants who received few \textit{likes} reported higher feelings of rejection (see table \ref{tab:manwhu} and figure \ref{fig:selfesteem}).


Participants receiving many \textit{likes} reported significantly lower levels of loneliness after the five days of social media interaction on our platform (Mdn =  24.0 , M =  25.67 , SD =  8.62) than before (Mdn= 26, M = 26.64 , SD = 8.04; Difference: \textit{Z}= -2.06, \textit{p}= .04*). In contrast, participants in the few \textit{likes} condition did not show any significant changes in regards to loneliness (\textit{Z}= -1.5 , \textit{p}= .14), although their average level of loneliness was smaller after the social media interaction (Mdn =  27.0 , M =  27.48 , SD =  8.0), than before (Mdn =  25.0 , M =  26.8 , SD =  8.38). 

\subsection{Self-Esteem}
When inspecting the situational self-esteem, participants receiving few \textit{likes} reported significantly lower levels of self-esteem during the social media interaction on Fakebook than participants receiving many \textit{likes} (see table \ref{tab:manwhu} and figure \ref{fig:selfesteem}). 
In comparison, the more stable, context-independet self-esteem measure was not affected in the same way by the social media interaction.
Participants in the many \textit{likes} condition reported the same level of self-esteem before (Mdn =  32.0 , M =  31.66 , SD =  5.27) and after the social media interaction (Mdn =  32.0 , M =  32.09 , SD =  4.92; Difference: \textit{Z}= -1.38 , \textit{p}= .16). Participants, in the few \textit{likes} condition, in contrast, on average, showed slightly lower scores of self-esteem after the treatment (Mdn =  33.0 , M =  31.93 , SD =  5.53) than before (Mdn =  32.0 , M =  31.16 , SD =  5.79), though not significant  (\textit{Z}= -1.94 , \textit{p}= .05). 

\subsection{Reasoning}
When asked about why they received this amount of \textit{likes} during the social media interaction, most participants (71\% in the many likes condition, 46\% in the no likes condition) stated a reason related to their post content, e.g. "Because my posts were [not] interesting/creative". 

This was followed by participants stating no reason (9\% many likes, 31\% few likes), e.g. "I have no idea", or reasons related to posting time (8\% many likes, 20\% few likes), e.g. "Because I published my posts early/late during the day".

16 participants (6\% many likes, 13\% few likes) also suspected reasons related to the study design, e.g. that the other participants were instructed to not like their posts or were not real persons. However, the simple suspicion did not affect how the participants reacted to receiving \textit{likes}/no \textit{likes}, as all of the following analyses were repeated with excluding participants' suspecting bot-behaviour, showing no different effects regarding situational emotions, social connectedness and self-esteem. 

Further, six participants stated reasons related to their person (5\% many likes, 2\% few likes), e.g. "Because I am cool" or "Because I am not a likeable person".
It should be emphasized here that participants who received many \textit{likes}, to a higher percentage stated reasons related to their post content and or to their person, while participants receiving no \textit{likes}, to a higher percentage attributed this to external reasons, such as posting time or study design.

\section{Discussion}
By conducting a social media experiment within a controlled environment, we could evaluate the effects of peer feedback on emotions, feelings of social connectedness and self-esteem. Based on a five-day intervention, receiving no positive online feedback did negatively impact the social media experience of users. In specific, it evoked negative emotions, such as stress, sadness and anxiety, left participants feeling rejected, less valued and less well received by their peers, and reduced the degree to which they felt like a person of worth during the online experience. It did not, however, impact the overall, context-independent self-esteem of users.

\subsection{Limitations and Future Work}
While the study has shown various implications of receiving online feedback, there are some questions which cannot be answered by design. Based on the current study, there are multiple explanations for the negative effects of lacking positive feedback. First of all, as suggested by the findings of \cite{Gallagher2017}, the negative impact might arise due to a mismatch between the amount of positive feedback one expected and the amount of feedback one received. Future work could evaluate this hypothesis by using a similar study design, which includes measuring the amount of expected feedback. In addition, future research could study how users develop feedback expectations, for example whether they take factors such as the amount of past feedback, the amount of feedback others receive or the perceived content quality into account.
Second, the effects might be induced by a social comparison. Thus, receiving few \textit{likes} in general may not be the problem, but receiving fewer \textit{likes} than others, could be. Future research should analyse, whether not displaying the \textit{like} counts of other users, but only showing the user's own \textit{like} count, could diminish the negative impacts of online ostracism.

In addition, the study only analyzed the effects of social media feedback received over the duration of five days (approx. 1,5 hours in total). During these days, users might have been exposed to other content, e.g. their regular social media activity, which could influence the study results. By utilizing an experimental design, which randomly assigned study participants to two different conditions, we hoped to control for these confounding variables. Moreover, the intervention might have been too short to generate significant changes in users' levels of context-independent self-esteem. Future research could use a similar design for a duration of several weeks, which measures the evolution of self-esteem during this time frame. Nevertheless, careful considerations need to be taken when conducting such studies, as they might impose negative long-term effects on users' mental health.

Even more the study only analyzed the effects of positive feedback and the lack thereof, without taking the effects of actual negative feedback into account. Future research could analyze how people react to negative social affordances, such as \textit{dislikes}, and negative comments.


\subsection{Contributions and Implications}
The prominent role social media plays in our everyday lives imposes new challenges for our society. Over the last decade, much research has analysed the positive and negative impact social media can have on users' mental health \cite{Braghieri2022, Sadagheyani2021}. However, the complex dynamics in this relationship are not well understood yet. Our work has contributed to this research in two ways.

First of all, we have revealed new insights into the relationship between social media use and mental health. Consistent with prior work \cite{Lee2020, Wolf2014}, the study demonstrated that people, who receive little positive feedback experience negative emotions, stress, and feelings of rejection on social media. As several studies have shown, these emotions and feelings can lead to severe mental health consequences when evoked on a regular basis, such as depression and anxiety \cite{Panak1992, McEwen2012}. At the same time, the study also revealed that social media use can help to reduce loneliness, given that users receive much positive online feedback. Thus, peer feedback plays a crucial role in influencing whether social media use impacts the user's mental health in a positive or negative way.

Interestingly, these findings did not seem to be influenced by whether participants believed the other users were bots. While the amount of bot suspecters was too small to generate any significant results, the mean values also indicated lower levels of stress and feelings of rejection, as well as higher levels of enjoyment, feelings of belongingness and appraisal, for people who received positive feedback from suspected bots in comparison to those who received few feedback from suspected bots. Specifically, one participant stated, "I suspect that the number of reactions is a targeted part of the study; Nevertheless, I feel valued with above-average likes on my posts". Future research could continue investigating this connection and evaluate whether \textit{likes} from accounts marked as bots help to reduce the negative impact of social media on mental health.

Secondly, we have advanced the methodological corpus of social media research by developing an open-source tool\footref{note:tool}, with which social scientists can design and conduct social media experiments. This tool can be used by researchers to study complex interactions and behavioral patterns in a controllable social media environment, and thus might help scientists to become independent from collaborations with social media companies.

\bibliographystyle{acm}
\bibliography{Bibliography}

\end{document}